\documentclass[12pt,preprint]{emulateapj}
\slugcomment{Submitted to {\it ApJL} on 30 Aug 2007; accepted 20 Sep
2007; published 10 Nov 2007}
\shortauthors{Looper et al.}
\shorttitle{DENIS 1245}

\begin{document}

\title{Discovery of an M9.5 Candidate Brown Dwarf in the TW Hydrae 
Association $-$ 
DENIS J124514.1$-$442907\footnote{Some of the data presented herein
were obtained at the W.M. Keck Observatory, which is operated as a 
scientific partnership among the California Institute of Technology, 
the University of California and the National Aeronautics and Space 
Administration. The Observatory was made possible by the generous
financial support of the W.M. Keck Foundation.  This 
paper includes data
gathered with the 6.5-m Magellan Telescopes located at Las Campanas
Observatory, Chile.}}

\author{Dagny L.\ Looper\altaffilmark{1,2},
Adam J.\ Burgasser\altaffilmark{3},
J.\ Davy Kirkpatrick\altaffilmark{4},
Brandon J.\ Swift\altaffilmark{1,5}}

\altaffiltext{1}{Institute for Astronomy, University of Hawaii, 2680
Woodlawn Drive, Honolulu, HI 96822}
\altaffiltext{2}{Visiting Astronomer at the Infrared Telescope
Facility, which is operated by
the University of Hawaii under Cooperative Agreement NCC 5-538 with
the National Aeronautics
and Space Administration, Office of Space Science, Planetary Astronomy Program}
\altaffiltext{3}{Massachusetts Institute of Technology, Kavli
Institute for Astrophysics and Space Research,
Building 37, Room 664B, 77 Massachusetts Avenue, Cambridge, MA 02139;
ajb@mit.edu}
\altaffiltext{4}{Infrared Processing and Analysis Center, M/S
100-22, California Institute of Technology, Pasadena, CA 91125}
\altaffiltext{5}{Currently at Steward Observatory, University of
Arizona, 933 N. Cherry Ave.,
Tucson, AZ 85721}

\begin{abstract}

We report the discovery of a fifth candidate substellar system in the
$\sim$5--10 Myr TW Hydrae association -- DENIS J124514.1$-$442907.   
This object has a
NIR spectrum remarkably similar to that of 2MASS J1139511$-$315921,
a known TW Hydrae brown dwarf, 
with low surface gravity features such as a triangular-shaped $H$ band, 
deep H$_2$O absorption, weak alkali lines, and weak 
hydride bands.  We find an optical spectral type of M9.5 and estimate 
a mass of $\lesssim$24 M$_{Jup}$, assuming an age of $\sim$5--10 Myr.  
While the measured proper motion for DENIS J124514.1$-$442907 
is inconclusive as a test for membership, its position in the sky is 
coincident with the TW Hydrae association.  A more accurate proper motion
measurement, higher resolution spectroscopy for radial velocity, 
and a parallax measurement are needed to derive the true
space motion and to confirm its membership.

\end{abstract}


\section{Introduction}

Although one of the nearest groups of newly formed stars, 
the TW Hydrae Association (TWA, \citealt{1997Sci...277...67K}; 
d$\sim$20--130 pc, age$\sim$5--10 Myr;
\citealt{2004ARA&A..42..685Z}) was only recently identified due
to its low spatial density over large spatial extent ($\alpha$: 10$^h$
to 14$^h$, $\delta$: $-$20$^\circ$ to $-$60$^\circ$).  
The association is composed of 25 known systems with 
early-type A to early-type M stellar primaries, 
with a high fraction of companions 
(0.84~$\pm$~0.22, \citealt{2003AJ....126.2009B}).  These objects
share similar kinematics ($\mu_{\alpha}$: $-$34 to $-$122
mas yr$^{-1}$, $\mu_{\delta}$: $-$10 to $-$43 mas yr$^{-1}$, and
V$_{rad}$: +6 to +17 km s$^{-1}$; \citealt{2003MNRAS.342..837R}), 
supporting their common motion through space.

In addition, four substellar systems in TWA have been identified -- 
two of which are not orbiting higher mass stars: 2MASSW 
J1139511$-$315921\footnote{After objects are introduced they are 
referred to by shortened versions of their names 
in the form [prefix + hhmm], where ``prefix'' is the survey
identifier name and ``hhmm'' is the truncated hour and minute portion of
the J2000 coordinates.} and 2MASSW J1207334$-$393254AB 
\citep{2002ApJ...575..484G}, and two of which are in systems with
stellar primaries: TWA 5B, which is a companion to TWA 5A 
\citep{1999ApJ...512L..69L,1999ApJ...512L..63W} and 
SSSPM J1102$-$3431, which might be a common proper motion 
companion to the star TW Hydrae \citep{2005A&A...430L..49S}.  
The substellar binary 
2MASS 1207AB has been the subject of intense study, revealing both 
a disk around the primary \citep{2004A&A...427..245S}, and possibly the
secondary \citep{2007ApJ...657.1064M}.  2MASS 1207A has also been
confirmed as an accretor by high resolution optical spectroscopy 
\citep{2003ApJ...593L.109M}.
The B component, a late-type L
dwarf, is believed to have a mass of $\sim$8 M$_{Jup}$ 
\citep{2004A&A...425L..29C,2005A&A...438L..25C,2007ApJ...657.1064M}.
The identification of additional low-mass members offers 
prototypes at a constrained age for atmospheric studies, testbeds for disk
studies at the low-mass end, an opportunity to examine low-mass star
formation processes in sparse associations, and excellent targets for
identifying planetary-mass companions.

In this Letter, we report the discovery of a fifth candidate brown dwarf
system in TWA, DENIS J124514.1$-$442907, identified 
serendipitously in the course of our spectroscopic follow-up of nearby,
late-type candidates (see \citealt{2007arXiv0710.1123B} for a description) 
in the Deep Near-Infrared Survey of the Southern Sky 
(DENIS; \citealt{1997Msngr..87...27E}).  We present the kinematics
of DENIS 1245 in $\S$2, spectroscopic observations in $\S$3, and
discuss its properties in $\S$4.

\section{Kinematics}

Using a linear fit to 
SuperCOSMOS Sky Survey (SSS; 
\citealt{2001MNRAS.326.1315H,2001MNRAS.326.1295H,2001MNRAS.326.1279H}), 
DENIS, and 2MASS time-ordered astrometry covering 4.1 yr (see Table 1), we found
a proper motion of $\mu_{\alpha}$~=~$-0\farcs$22~$\pm$~0$\farcs$07
yr$^{-1}$ and $\mu_{\delta}$~=~+0$\farcs$02~$\pm$~0$\farcs$10
(not including
parallactic reflex motion; see Table 2).  The motion
errors include an estimated uncertainty of 0$\farcs$3 in
both right ascension and declination for all three epochs.

In Table 3, we list the measured motion 
for DENIS 1245 and the four known TWA brown dwarf systems.  
While $\mu_{\alpha}$ 
of DENIS 1245 is in the westerly direction, it is 
2.1$\sigma$ away from the 
mean and 1.4$\sigma$ away from the edge of the range in 
$\mu_{\alpha}$ of TWA stellar systems 
\citep{2003MNRAS.342..837R}.  
On the other hand, $\mu_{\delta}$ of DENIS 1245 is in
the northerly direction, but its large error places it less than 
0.5$\sigma$ away from the mean $\mu_{\delta}$ of TWA stellar systems 
\citep{2003MNRAS.342..837R}.  
Clearly, at least one more, precise astrometric measurement is
needed to better constrain the motion of DENIS 1245 and to test its
membership in TWA.

\section{Observations}

\subsection{NIR Spectroscopy}

We observed 2MASS 1139, 2MASS 1207A\footnote{Although the B component of 
2MASS 1207AB is unresolved from the A component in our observations, 
it makes a negligible contribution
($\Delta~J_{B-A}\sim$~7; \citealt{2007ApJ...657.1064M}) to the spectrum
of 2MASS 1207A.  We therefore will refer to the spectroscopy of 2MASS
1207AB as that for 2MASS 1207A alone.}, and DENIS 1245 on three 
consecutive nights: 2007 Mar 16, 17, \& 18 (UT), respectively, using 
the Infrared Telescope Facility's (IRTF) SpeX spectrograph
\citep{2003PASP..115..362R}.    
Skies were clear or contained only light cirrus 
with $\sim1\arcsec$ seeing at $K$.
We used the prism mode
with the 0$\farcs5\times15\arcsec$ slit (rotated to the parallactic angle), 
yielding a resolution of
R~=~$\lambda$~/~$\Delta\lambda~\approx$~150.  
2MASS 1139 was observed at an airmass of 1.9 in four AB nodding cycles 
with 120 seconds of exposure time per nod position.   
2MASS 1207A was observed at an airmass of 2.0 with three AB nodding
cycles with 120 seconds of exposure time per nod position.  
DENIS 1245 was observed at an airmass of 2.4 with two cycles AB nodding
cycles with 120 seconds of exposure time per nod position.
For flux calibration and telluric correction, the A0 V stars 
HD 100330 for 2MASS 1207A and HD 103870 for 2MASS 1139 and DENIS 1245 
were observed at a similar airmass and time as the targets, followed by 
flat-field and argon arc lamp calibration frames.  
The data were reduced in standard fashion using 
the Spextool package version 3.2 
\citep{2004PASP..116..362C,2003PASP..115..389V}.

The reduced NIR spectra are shown in
Figure 1 along with the M8 standard VB 10 (Kirkpatrick et al., in prep) 
and the L0 standard 2MASS J03454316+2540233 (Cruz et al., in prep), 
also taken in SpeX prism mode with
identical settings.    
DENIS 1245, like 2MASS 1139 and 2MASS 1207A, shows
several hallmarks of low surface gravity, such as a triangular-shaped
H-band, deep H$_2$O absorption at 0.89--0.99 $\mu$m, 1.30--1.51 $\mu$m 
\& 1.75--2.05 $\mu$m, stronger VO absorption, 
and weaker alkali absorption in comparison to field
standards.  Also, the FeH Wing-Ford band at 0.99 $\mu$m in the spectra
of 2MASS 1139, 2MASS 1207A and DENIS 1245 is very weak in
comparison to the field M8 and L0 standards.  

Determining NIR spectral types of young brown
dwarfs is difficult, given that the deep H$_2$O absorption at 
1.30--1.51 \& 1.75--2.05 $\mu$m 
tends to lead to later spectral typing in the NIR than in the
optical.  To this end, \cite{2007ApJ...657..511A} have created an H$_2$O
index (1.49--1.59 $\mu$m), 
largely insensitive to surface gravity, that can be used
to spectrally classify M5--L5 dwarfs in the NIR.  We use this index
to classify 2MASS 1207A, 2MASS 1139 and DENIS 1245 in the NIR.  
Overall, the NIR 
morphology of DENIS 1245 is nearly identical to that of 2MASS 1139,
suggesting that the two have similar surface gravities.  

\subsection{Optical Spectroscopy}

Red optical spectral data for DENIS 1245 were obtained on
2007 May 9 (UT) in clear conditions
using the Low Dispersion Survey Spectrograph (LDSS-3; \citealt{all94})
mounted on the Magellan 6.5m Clay Telescope.  Data were obtained using the
VPH-red grism (660 lines/mm) with a 0$\farcs$75
(4-pixel) wide longslit aligned to the parallactic angle,
providing 6050--10500 {\AA} spectra
at a resolution $\lambda$~/~$\Delta\lambda$ $\approx$ 1800.  
The OG590 longpass filter
was used to eliminate second order light shortward of 6000 {\AA}.
A single exposure of 1500~s was obtained at an airmass of 1.08,
followed immediately by observation of the G2~V star HD~114853
for telluric absorption correction.  Dispersion and pixel
response calibration were determined by HeNeAr arc lamp and flat-field
quartz lamp exposures obtained immediately after observation
of both sources.  The data reduction, carried out in the IRAF\footnote{IRAF
is distributed by the National Optical
Astronomy Observatories, which are operated by the Association of
Universities for Research in Astronomy, Inc., under cooperative
agreement with the National Science Foundation.}
environment, is described in detail in \citet{bur07}.

The reduced red optical spectra of 2MASS 1139, 2MASS 1207A
(R$\sim$1200, Kirkpatrick et al., in prep) 
and DENIS 1245 (R$\sim$1800) are presented in Figure 2,
along with the M8.5 TWA brown dwarf SSSPM 1102 (R$\sim$600;
\citealt{2005A&A...430L..49S}), the M8 comparison star 2MASS J14342644+1940499 
and the L0 standard 2MASS 0345 
(R$\sim$1200; \citealt{1999ApJ...519..802K}).  Several features in the
spectra of 2MASS 1139, 2MASS 1207A and SSSPM 1102 
supporting
their low surface gravity are also seen in the spectrum of
DENIS 1245.  In particular, stronger VO and TiO absorption and weaker CaH and K I
absorption in comparison to the field standards.  There is also no
discernible FeH absorption at 8680--8820 $\AA$ in the spectra of the
three TWA brown dwarf systems and DENIS 1245.  

The TWA brown dwarf systems and DENIS 1245 
show prominent H$\alpha$ emission at 6563 $\AA$ (see bottom
insert of Figure 2).  We have measured the EW of this feature in all
four objects (see Figure 2) using the SPLOT package in IRAF.  
The uncertainties in these
measurements were estimated using the standard deviation of five
measurements taken for each emission profile, 
added in quadrature to the
average of five measurements of a typical noise spike in each spectrum.
SSSPM 1102, which shows an asymmetric H$\alpha$ profile, has the
strongest measured emission, with an EW of
64~$\pm$~3 $\AA$\footnote{Note that \cite{2005A&A...430L..49S} measure 
an EW of 10 $\AA$ using this same spectrum.  This value was in fact a
FWHM measurement and not an EW as stated (personal communication, R.\ Scholz).}
DENIS 1245 has the weakest measured EW of 15~$\pm$~3 $\AA$. 
It should be noted that the strength and morphology of 
H$\alpha$ emission in the spectrum of 2MASS 1207A changes dramatically over time 
\citep{2005ApJ...629L..41S,2006ApJ...638.1056S}.  Future observations
will allow us to determine whether variable emission is also present in
DENIS 1245.

Since the three TWA substellar systems and DENIS 1245 become much redder than the
field standards at wavelengths greater than 
$\sim$8500 $\AA$, we have spectrally classified
DENIS 1245 by its morphology at wavelengths shorter than this.
DENIS 1245 appears to be later in spectral type than all three TWA brown
dwarf systems.  We assign an optical spectral type of M9.5 based on
visual comparisons
to late-type M and early-type L dwarf optical standards (see Figure 2).

\section{Discussion}

In comparison to the four known TWA brown dwarf systems, DENIS 1245 is the
faintest in $J$ and also the reddest in 
$I-J$.  These differences likely arise from DENIS 1245 being more distant and
possibly cooler.  
Determining spectrophotometric distance estimates for young brown
dwarfs is not straightforward, as well-established absolute
magnitude/spectral relations based on evolved field dwarfs (e.g., Cruz
et al. 2003) can significantly underestimate the distances of these
overluminous, low surface gravity objects.  However,  \cite{gizis2007}
have recently measured the parallax of 2MASS 1207A, and we can
use the absolute magnitude of this source as a proxy for DENIS 1245.
Based on its $J$ magnitude and assuming similar brightness, we
find d$\sim$110 pc for DENIS 1245.  However, 
this estimate is likely an upper
limit since DENIS 1245 has an optical spectral type 1.5 subtypes later
than that of 2MASS 1207A.  \cite{2003AJ....126.2421C} find a
$\Delta$M$_{J}$~=~0.44 mag difference between optical spectral
types M8 and M9.5.  If
this difference were then applied to the $J$ magnitude of DENIS
1245, we would then find an estimated distance of $\sim$90 pc.
So we assume a rough distance estimate of 100 pc for DENIS 1245.

Since the three other TWA brown dwarf systems have spectral types similar
to that of 2MASS 1207A, we can use the parallax
measurement for 2MASS 1207A and their $J$ magnitudes 
to estimate distances to each of these objects (see Table 3).  
\cite{2005A&A...430L..49S} have suggested that SSSPM 1102 is a
widely separated companion to the star TW Hydrae.  Our distance estimate to SSSPM
1102 ($\sim$55 pc) agrees well with the distance to TW Hydrae ($56~\pm~7$ pc;
\citealt{1997A&A...323L..49P}).

We note that if DENIS 1245 is indeed a TWA member, its M9.5 optical
spectral type implies that it is cooler and less massive than the M8
optical brown dwarf TWA 1207A (T$_{eff,A}\approx2550~\pm~150$ K,
M$_A\approx$~24~$\pm$~6 M$_{Jup}$; \citealt{2007ApJ...657.1064M}).  This
upper limit on mass (M~$\lesssim$~24 M$_{Jup}$) places DENIS 1245 firmly
in the substellar regime.  

DENIS 1245 is only the fifth substellar system to be identified within 
the TW Hydrae Association.  It was not uncovered by
\cite{2002ApJ...575..484G} as it
is 1 mag fainter in 2MASS $K_s$ than the faintest objects identified in that
study with similar J$-K_s$ colors.  
Our result therefore suggests that more distant and later-type
substellar members of TWA may remain to be identified.  Nevertheless,
DENIS 1245 requires further high resolution spectroscopic observations
and a more accurate proper motion determination to verify that it
shares a common space motion with TWA.  These observations, coupled
with mid-infrared photometry, can also test whether this source, like
2MASS 1207A (and possibly B), harbors a disk.  
Lastly, as a very young, late-type brown dwarf, DENIS 1245
presents an excellent target for high resolution imaging for
planetary/substellar companions.

\section{Acknowledgments}

We thank our referee S$.$ Mohanty for suggestions which significantly
improved the manuscript.  
DLL thanks M$.$ Pitts for comments on the manuscript, M$.$ Cushing for
useful discussions, J$.$ Gizis for providing an early copy of his
manuscript on the parallax for 2MASS 1207A, 
and R$.$ Scholz and K$.$ Cruz for providing spectra.  We also 
thank our telescope operators B$.$ Golisch, D$.$ Griep and   
H$.$ Nu\~{n}ez, and instrument specialist J$.$ Rayner
for their assistance.  
This publication includes data from the DENIS 
project, which has been partly funded by the SCIENCE and the HCM plans of
the European Commission under grants CT920791 and CT940627.
This publication makes use of data products 
from the Two Micron All Sky 
Survey (2MASS), which is a joint project of the University of
Massachusetts and the Infrared 
Processing and Analysis Center/California Institute of Technology, 
funded by the National Aeronautics and Space Administration and the 
National Science Foundation.  
This research has made use of the NASA/IPAC Infrared Science Archive  
(IRSA), which is operated by the Jet Propulsion Laboratory,  
California Institute of Technology, under contract with the National  
Aeronautics and Space Administration. 
This research used the facilities of the 
Canadian Astronomy Data Centre operated by the National Research Council 
of Canada with the support of the Canadian Space Agency.  

\textbf{Facilities Used:} Keck I 10.0-m/LRIS, Magellan 6.5-m/LDSS-3 \& 
IRTF 3.0-m/SpeX.

\clearpage

\begin{deluxetable}{llll} 
\tablewidth{3in}
\tablenum{1}
\tablecaption{Astrometry for DENIS J124514.1$-$442907}
\tablehead{ 
\colhead{$\alpha$\tablenotemark{a}} & \colhead{$\delta$\tablenotemark{a}} & 
\colhead{Epoch} & \colhead{Catalog}}
\startdata
12 45 14.25 & $-$44 29 07.8 & 27 Mar 1996 & UKST; SSS \\ 
12 45 14.20 & $-$44 29 07.7 & 03 May 1999 & DENIS \\
12 45 14.16 & $-$44 29 07.7 & 27 Apr 2000 & 2MASS \\
\enddata
\tablenotetext{a}{Equinox J2000.}
\end{deluxetable}

\begin{deluxetable}{lll} 
\tablewidth{3in}
\tablenum{2}
\tablecaption{Properties of DENIS J124514.1$-$442907}
\tablehead{ 
\colhead{Parameter} & \colhead{Value} & \colhead{Ref}}
\startdata
$\alpha$ (J2000)\tablenotemark{a} & 12 45 14.20 & 1 \\
$\delta$ (J2000)\tablenotemark{a} & $-$44 29 07.7 & 1 \\
$\mu_{\alpha}$ & 0$\farcs$22~$\pm$~0$\farcs$07 yr$^{-1}$ & 2 \\
$\mu_{\delta}$ & 0$\farcs$02~$\pm$~0$\farcs$10 yr$^{-1}$ & 2 \\
d$_{est}$ & $\sim$100 pc & 2 \\
Optical SpT & M9.5 & 2 \\
NIR SpT & M9 pec & 2 \\
Age\tablenotemark{b} & $\sim$5--10 Myr & 2 \\
Mass\tablenotemark{c} & $\lesssim$24 M$_{Jup}$ & 2 \\
$I_N$ & 17.94 mag & 3 \\
$I$ & 18.0~$\pm$~0.2 mag & 1 \\
$J$ & 14.52~$\pm$~0.03 mag & 4 \\
$H$ & 13.80~$\pm$~0.03 mag & 4 \\
$K_s$ & 13.37~$\pm$~0.04 mag & 4 \\
$I-J$ & 3.5~$\pm$~0.2 mag & 1,4 \\
$J-H$ & 0.72~$\pm$~0.04 mag & 4 \\
$H-K_s$ & 0.43~$\pm$~0.05 mag & 4 \\
$J-K_s$ & 1.15~$\pm$~0.05 mag & 4 \\
\enddata
\tablenotetext{a}{Coordinates are from the DENIS 3rd Release Catalog at
epoch 03 May 1999.}
\tablenotetext{b}{Based on age estimates of TWA  
\citep{2005MNRAS.362.1109M,2006AJ....131.2609D,2006A&A...459..511B}.}
\tablenotetext{c}{This upper mass limit is based on the mass estimate
for 2MASS 1207A \citep{2007ApJ...657.1064M}.}
\tablecomments{References: (1) DENIS (\citealt{1997Msngr..87...27E}), 
(2) this paper, 
(3) SSS (\citealt{2001MNRAS.326.1315H,2001MNRAS.326.1295H,2001MNRAS.326.1279H}), \& 
(4) 2MASS (\citealt{2006AJ....131.1163S}).}
\end{deluxetable}

\begin{deluxetable}{lccccccccc} 
\tabletypesize{\scriptsize}
\tablewidth{6.5in}
\tablenum{3}
\tablecaption{Spectral Types \& Kinematics of TWA Brown Dwarfs}
\tablehead{ 
\colhead{Object} & \colhead{Opt SpT} & \colhead{Ref.} & 
\colhead{NIR SpT\tablenotemark{a}} & \colhead{H$_2$O Index} & 
\colhead{$\mu_{\alpha}$ [mas/yr]} & 
\colhead{$\mu_{\delta}$ [mas/yr]} & \colhead{Ref.} & 
  \colhead{Dist} & \colhead{Ref.}}
\startdata
SSSPM 1102 & M8.5 & 1 & \nodata & \nodata & $-82~\pm~12$ &
  $-12~\pm~6$ & 1 & $56~\pm~7$ & 1 \\
TWA 5B & \nodata & \nodata & M8--M8.5\tablenotemark{b}  & \nodata & 
  $-86~\pm~8$ & $-21~\pm~8$ & 4 & $\sim$45 & 3 \\
2MASS 1139 & M8 & 2 & M9pec\tablenotemark{c} & 1.13 & $-93~\pm~5$ &
  $-31~\pm~10$ & 1 & $\sim$47 & 3 \\
2MASS 1207A & M8 & 2 & M8pec\tablenotemark{c} & 1.08 & $-63~\pm~3$ &
  $-23~\pm~2$ & 5 & $54~\pm~3$ & 5 \\
DENIS 1245 & M9.5 & 3 & M9pec\tablenotemark{c} & 1.14 & $-220~\pm~70$ &
  $+20~\pm~100$ & 3 & $\sim$100 & 3 \\
\enddata
\tablenotetext{a}{Derived from the H$_2$O index
  \citep{2007ApJ...657..511A}, unless otherwise noted.}
\tablenotetext{b}{NIR photometric spectral type from \cite{1999ApJ...512L..69L}}.
\tablenotetext{c}{The label ``peculiar'' has been appended to the NIR
spectral type of these three objects based on their triangular-shaped
H-band and other low surface gravity features (see \citealt{2006ApJ...639.1120K}).}
\tablecomments{Optical Spectral Type, Kinematic and Distance References
  are (1) \cite{2005A&A...430L..49S}, (2) \cite{2002ApJ...575..484G}, 
  (3) this paper, (4) \cite{1999ApJ...512L..69L}, \& (5) \cite{gizis2007}.}
\end{deluxetable}

\clearpage


\begin{figure}
\epsscale{0.8}
\plotone{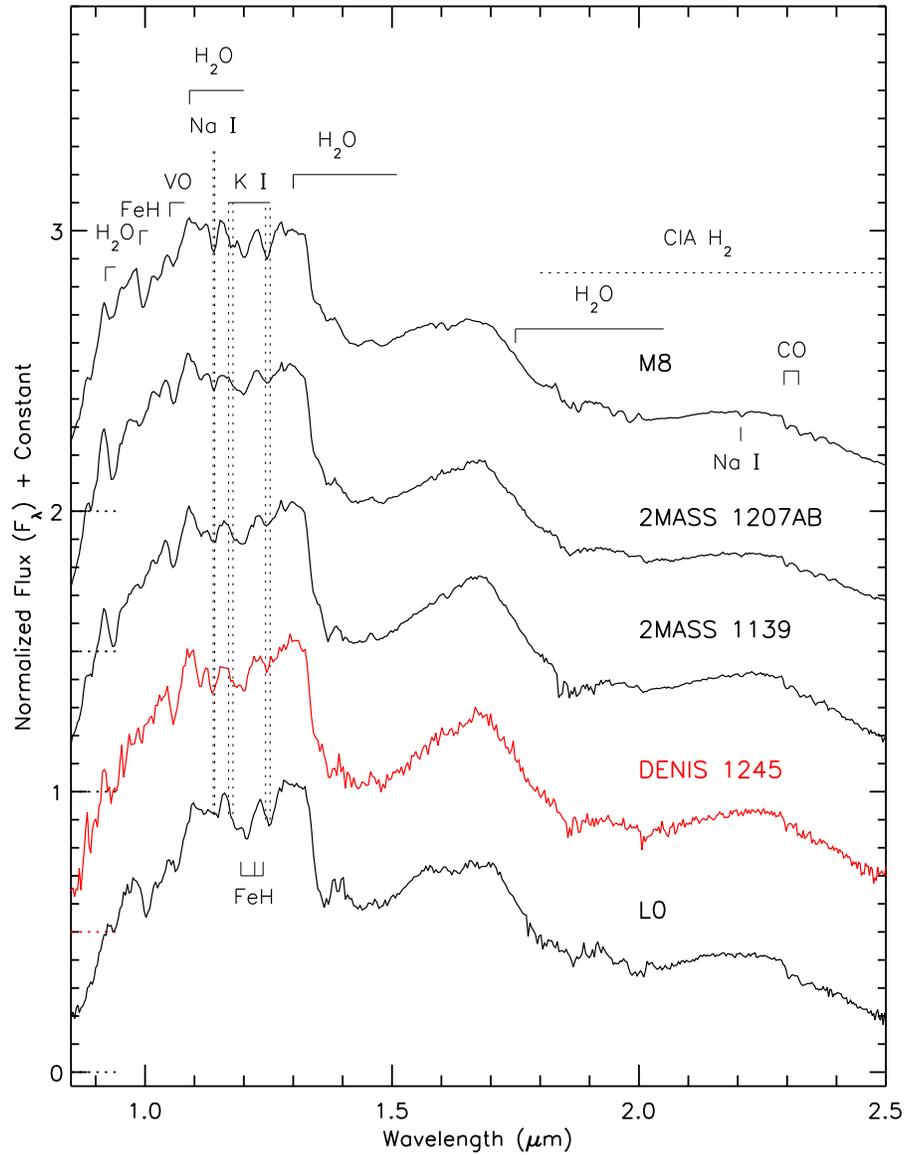}
\caption{NIR spectra of DENIS 1245 (red, M9pec), 2MASS 1207A (M8pec) and 2MASS
1139 (M9pec) taken with IRTF/SpeX in prism mode (R$\sim$150) shown in
comparison to the M8 spectral standard VB 10 and the 
L0 spectral standard 2MASS 0345 taken with an identical set-up.
All spectra
have been normalized at 1.27 $\mu$m and are offset by half-integer values 
(dotted lines show zero levels) for clarity.  Major
atomic and molecular features have been labeled.  
\label{fig1}}
\end{figure}

\begin{figure}
\epsscale{0.8}
\plotone{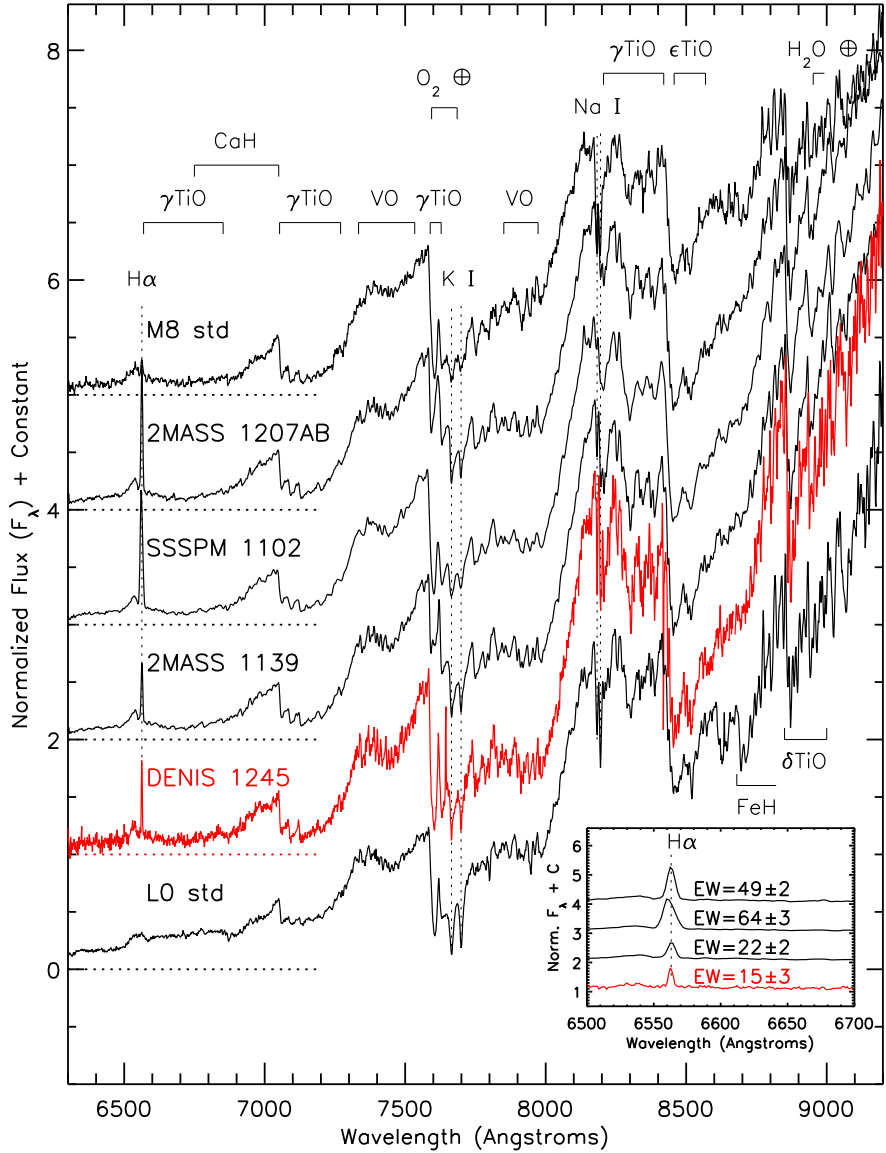}
\caption{The red optical spectra of DENIS 1245 (red, M9.5; R$\sim$1800),
2MASS 1207A and 2MASS 1139 (M8; R$\sim$1200).  
Shown for comparison are the M8
star 2MASS 1434, the L0 standard 2MASS 0345 (R$\sim$1200),  
and SSSPM 1102 (M8.5; R$\sim$600).  2MASS
1207AB, SSSPM 1102 and 2MASS 1139 are all known brown dwarf systems of
TWA.  All spectra have been normalized at 7500 $\AA$ and are offset by
integer values (dotted lines show zero levels) for clarity.  Major
atomic and molecular features have been labeled.  The bottom right
insert shows the detection of H$\alpha$ emission in the TWA brown
dwarfs and their measured equivalent widths (EW), following the same
order as the main figure.  
\label{fig2}}
\end{figure}

 \end{document}